FS%% Beginning of file 'sample.tex'
\documentclass[12pt,preprint]{aastex}

\slugcomment{Accepted: Astrophysical Journal, December 2008}

\shorttitle{Cepheid Distance to IC~1613}
\shortauthors{Freedman {\it et al.}}

\begin{document}


\title{The Cepheid Period-Luminosity Relation (The Leavitt Law) at
Mid-Infrared Wavelengths: IV. Cepheids in IC~1613}


\author{\bf Wendy L. Freedman, Jane Rigby, Barry F. Madore  \\ S. E. Persson,
Laura Sturch \& Violet Mager} \affil{Observatories of the Carnegie
Institution of Washington \\ 813 Santa Barbara St., Pasadena, CA
~~91101} \email{wendy@ociw.edu, jrigby@ociw.edu, barry@ociw.edu,
persson@ociw.edu, lsturch@ociw.edu, vmager@ociw.edu}

%% are identified by the \altaffilmark after each name.  Specify  alternate


\begin{abstract}
We present mid-infrared Period-Luminosity relations for Cepheids in
the Local Group galaxy IC~1613.  Using archival IRAC imaging
data from Spitzer we were able to measure single-epoch magnitudes for
five, 7 to 50-day, Cepheids at 3.6 and 4.5$\mu$m. 
When fit to the calibrating relations, measured for the Large
Magellanic Cloud Cepheids, the data give apparent distance moduli of
24.29$\pm$0.07 and 24.28$\pm$0.07 at 3.6 and 4.5$\mu$m, respectively. A
multi-wavelength fit to previously published BVRIJHK apparent moduli
and the two IRAC moduli gives a true distance modulus of 24.27 $\pm$
0.02~mag with E(B-V) = 0.08~mag, and a corresponding metric distance
of 715 kpc. Given that these results are based on single-phase
observations derived from exposures having total integration times of
only 1,000~sec/pixel we suggest that Cepheids out to about 2~Mpc are
accessible to Spitzer with modest integration times during
its warm mission. We identify the main limiting factor to this method
to be crowding/contamination induced by the ubiquitous population of
infrared-bright AGB stars.

\end{abstract}

\vfill\eject
\section{Introduction}

The Local Group galaxy, IC~1613 is a highly-resolved dwarf of type
IB(s)m (de Vaucouleurs et al. 1991).  IC~1613 is rich in gas and is
actively forming stars; as a consequence it also contains many
Classical (Population I) Cepheid variables. These stars have been
the subject of a variety of multi-wavelength distance determinations,
which place the host galaxy at a distance of about 700~kpc, comparable
to that of M31, the Andromeda nebula.  Consistent with its high
Galactic latitude, the foreground Galactic extinction towards IC~1613
appears to be low, with estimates ranging from E(B-V) = 0.025~mag
(Schlegel et al. 1998) to E(B-V) = 0.005~mag (Burstein \& Heiles
1982).

Cepheids in IC 1613 were first discovered by Hubble, Mayall and Baade
in the 1930s (as reported by Sandage 1971), but it was not until about
40 years later that Sandage published Baade's data on that galaxy. Of
the 59 variable stars reported at least thirty-seven were considered to
be bona fide Cepheids. Only B-band photographic photometry was
available at that time. Carlson \& Sandage (1990) updated the periods
and the B-band light curves for 16 of the faintest Cepheids in
IC~1613, extending the PL relation down to about 2 days. The first
multi-wavelength $BVRI$ CCD observations of Cepheids in IC~1613 were
made by Freedman (1988), allowing a simultaneous fit 
for the reddening and true distance modulus which were determined to be
E(B-V) = 0.04$\pm$0.04~mag and $\mu_o$ = 24.30$\pm$0.10~mag,
respectively.  In the meantime more recent optical studies of Cepheids
in IC~1613 include the unfiltered (``white light'') CCD surveys by
Antonello et al. (1999, 2000), Mantegazza, et al. (2001), and the VI
CCD monitoring project of Udalski et al. (2001) which raised the
numbers of known Cepheids in IC~1613 to at least 138), and the BVRI
sparse-sampling, follow-up study of Antonello et al. (2006) which
concluded that E(B-V) = 0.07$\pm$0.08~mag and $\mu_o$ =
24.23$\pm$0.20~mag.  Near infrared H-band observations of 10 Cepheids
in IC~1613 were first made by McAlary, Madore \& Davis (1984) giving
$\mu_o$ = 24.31$\pm$0.12~mag, and these were more recently
complemented by a major survey of 29 Cepheids at J and K wavelengths
published by Pietrzynski et al.~(2006).  They derived a reddening of
E(B-V) = 0.09$\pm$0.02~mag and a true distance modulus of $\mu_o$ =
24.29$\pm$0.04~mag.

\section{IRAC Observations and Mid-Infrared Period-Luminosity Relations}

As part of Guaranteed Time observations Gehrz (PID: 128) obtained a
series of observations of IC~1613 resulting in an average integration
time of about 16~min/pixel for all four mid-IR channels. Six uncrowded
Cepheids were measured at the shortest two wavelengths. All but two of
the Cepheids were either undetected or confused at 8.0$\mu$m. No
Cepheids in IC 1613 were confidently measured in IC 1613 at
5.8$\mu$m.  The very long-period (146 day) Cepheid IC~1613:[S71] V22
was measured in all three bands; however its period puts it beyond the
limits of our standard calibration and it was not used in this
determination of a distance to IC~1613. We do note, however, that V22
does conform to the trend noticed for these very long-period Cepheids
(known as Leavitt Variables, Grieve, Madore \& Welch 1985) in that
they are subluminous with respect to a linear extrapolation of
period-luminosity relations calibrated on shorter-period (10 to
60-day) Cepheids.

Fluxed mosaics, known as ``post-BCD mosaics'', were downloaded from
the Spitzer archive. Aperture photometry was done using DAOPHOT with an
R $=$ 2 native pixel source aperture, and an R $=$ 2.6 pixel annular
sky aperture. Aperture corrections were taken from Table 5.7 of the
{\it IRAC Data Handbook}. No color corrections were applied.  The
observations were reduced using DAOPHOT psf fitting techniques in
exactly the same way as the IRAC observations of Cepheids in NGC~6822
(Madore et al. 2009b). The reader is referred to that paper for
details.  The final sample of Cepheids and their psf magnitudes are
given in Table 1.  The flux densities, converted to the IRAC magnitude
system use the zero-points of Reach et al.  (2005) which correspond to
the SAGE zero points used for the LMC calibration. Errors on the
individual data points are those returned by DAOPHOT and are based on
photon statistics alone.

\includegraphics [width=10cm, angle=270] {ic1613_irac.ps} 
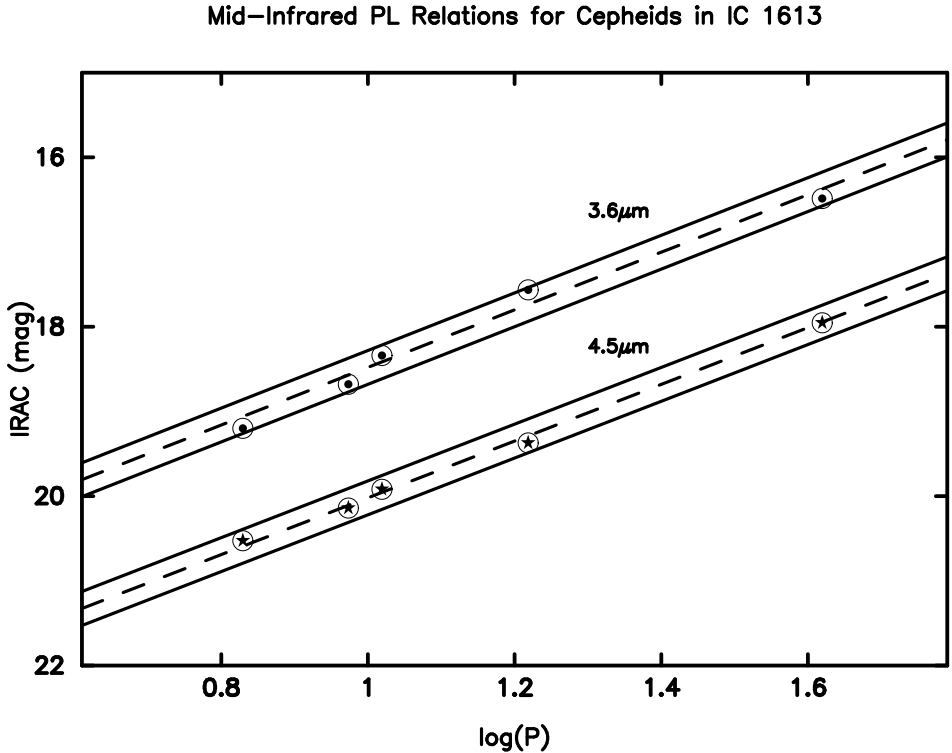
\par\noindent \bf Fig.~1 \rm -- Mid-Infrared Period-Luminosity
relations for Cepheids in IC~1613. Dashed lines are fits to the
fiducial relations of Madore et al. (2009a) based on LMC data and an
adopted true distance to the LMC of 18.50~mag. The flanking solid
lines are offset from the fit by $\pm$0.20~mag and appear to represent
the full width of each of the relations.

\section{Multi-Wavelength Solutions}

Table 2  contains apparent distance moduli to IC~1613 based on a
variety of published Cepheid samples measured at optical through near-infrared
wavelengths.  These moduli are plotted as a function of inverse
wavelength in Figure 2. The near flatness of the run of apparent
moduli with inverse wavelength immediately attests to the fact that
the total line-of-sight reddening towards IC~1613 and its Cepheids is
indeed small.  The plotted fit to a standard Galactic extinction
curve (Cardelli et al. 1989) is scaled to a color excess of E(B-V) =
0.08~mag, and a true distance modulus of 24.27$\pm$0.02~mag, which
corresponds to a metric distance of 715$\pm$7~kpc.

\includegraphics [width=10cm, angle=270] {ic1613_multi.ps}
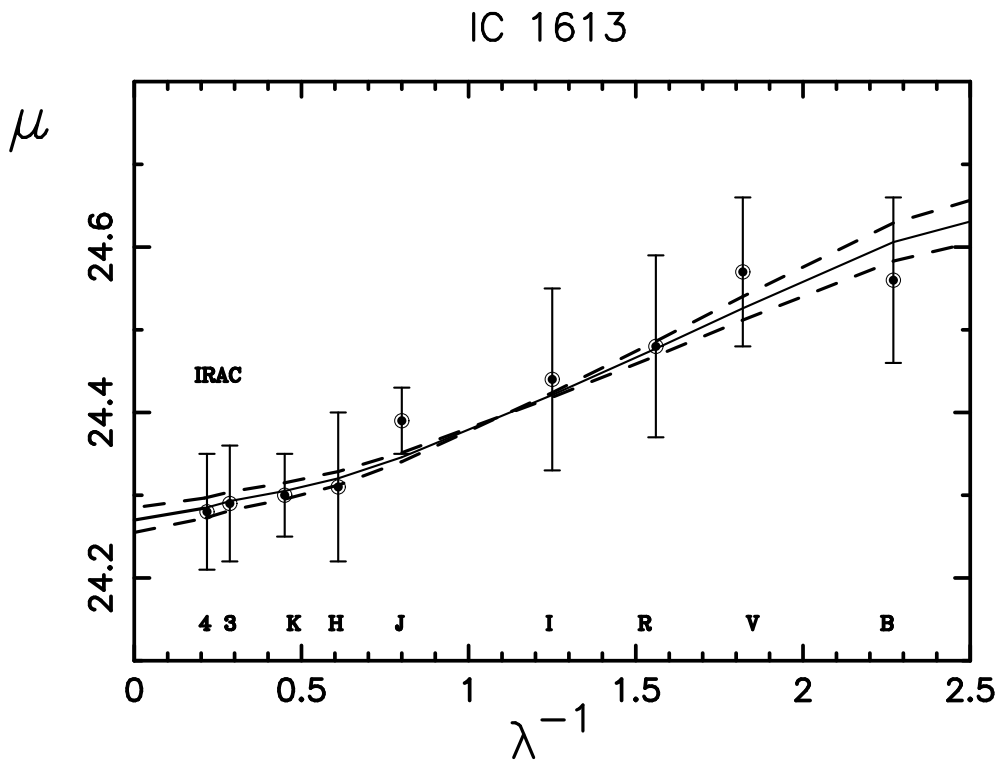
\par\noindent \bf Fig.~2 \rm -- A plot of apparent distance moduli as
a function of inverse wavelength. Data points are individual Cepheid
distance moduli. The solid line is a scaled and shifted Galactic
extinction curve whose slope is the extinction and whose intercept is
the true distance modulus. Individual bands contributing to the
solution are given just above the bottom axis.

We note here that these observations were reduced in exactly the same
manner as the IRAC observations of Cepheids in NGC~6822 (Madore et al.
2009b), although there are differences in the post-processing of the
mosaics, with the SINGS images of NGC~6822 being enhanced over the
standard post-bcd mosaics used here.  In the NGC~6822 paper we noted a
discrepancy between the mid-infrared distance moduli and the two
near-infrared (J \& K) moduli from (Gieren et al.~2006).  No such
discrepancy is seen in a comparison of these IRAC observations and the
near-infrared (again J \& K) moduli for IC~1613, this time published
by Pietrzynski et al.~(2006). As such we are still no closer to
understanding the differences seen in the NGC~6822 near-infrared data
sets.

\section{Limits on IRAC}

Given the sensitivity of IRAC, especially for the shortest wavelength
bands, modest integration times (hours) could, in principle, allow one
to press the application of mid-infrared period-luminosity relations
in determining distances out to several megaparsecs. Scaling from the
data on IC~1613, similar Cepheids at a distance of 2~Mpc could be
measured to the same signal-to-noise in about three hours. 

However, sensitivity is not the limiting factor: crowding is. The
mid-infrared studies of the stellar populations in WLM and IC~1613 by
Jackson et al. (2007a,b) have convincingly shown that the vast
majority of stars resolved at the brightest magnitudes (i.e., above
the tip of the red giant branch, M$_{3.6}$ = -6.0~mag and brighter)
are IR-AGB stars, a substantial component of which (40-45\%) are not
detected in the optical. From the data presented in Figure 5 of
Jackson et al. (2007a) it is possible to estimate the areal density of
these stars across the face of IC~1613. From that we deduce that the
mean separation of stars brighter than -6~mag at 3.6$\mu$m is
18~arcsec. Translated into practical terms this means that for a
galaxy 10 times further than IC~1613 (i.e. 7-8~Mpc) the average
separation of bright sources would be less than 2~arcsec, which is now
comparable to the resolution of IRAC. Even at the distance of IC~1613
itself we found that upwards of half of the known Cepheids in this
galaxy were visibly contaminated.

Thus we estimate that care must be taken in using IRAC to measure
Cepheids in galaxies another factor of two or three further in
distance than IC~1613.  More than half of the Cepheids will be crowded
but for a given pointing the total number of Cepheids that fit in the
detector's field of view will also go up so that the absolute yield of
uncontaminated Cepheids might be maintained at this limit. Pressing
this method beyond 2~Mpc is best left for JWST; but everything inside
of that sphere is plausibly within reach.

\section{Conclusions}

Using short-exposure, archival data from Spitzer we have demonstrated
the feasibility of measuring mid-infrared magnitudes for Cepheids out
to one of the most distant galaxies in the Local Group. Exposures a
factor of 10 or more longer would only amount to a couple of hours and
are not unreasonably long. These could allow one to press observations
of Cepheids using IRAC on Spitzer out to 2~Mpc and into the nearby
Hubble flow. The main limiting factor is crowding and contamination of
the Cepheid photometry by the rise of infrared-bright AGB and extended
AGB.

The two short-wavelength IRAC distance moduli to IC~1613 give
individual distance moduli of $\mu_{3.6}$ = 24.29$\pm$0.07 and
$\mu_{4.5}$ = 24.28$\pm$0.07~mag. When combined with optical (BVRI)
and near-infrared (JHK) data we derive a true distance modulus to
IC~1613 of $\mu_o$ = 24.27$\pm$0.02~mag (715~kpc) and a reddening of
E(B-V) = 0.08~mag. The uncorrected IRAC observations alone are each
within $\pm$0.02~mag of the finally adopted true modulus of IC~1613.

\vfill\eject
\noindent
\centerline{\bf References \rm}
\vskip 0.1cm
\vskip 0.1cm

\par\noindent Antonello, E., Fossati, L., Fugazza, D., Mantegazza, L.,
\& Gieren, W. 2006, \aap, 445, 901

\par\noindent Antonello, E., Fugazza, D., Mantegazza, L., Bossi, M.,
\& Covino, S. 2000, \aap, 363, 29

\par\noindent Antonello, E.,  Mantegazza, L., Fugazza, D., Bossi, M.,
\& Covino, S. 1999, \aap, 349, 55

\par\noindent Antonello, E., Mantegazza, L., Fugazza, D., \& Bossi,
M. 1999, \aap, 350, 797

\par\noindent Burstein, D., \& Heiles, C. 1982, \aj, 87, 1165

\par\noindent Cardelli, J.A., Clayton, G.C., \& Mathis, J.S. 1989,
\apj, 345, 245

\par\noindent
Carlson, G., \& Sandage, A.R. 1990, \apj, 352, 587

\par\noindent de Vaucouleurs, G., de Vaucouleurs, A., Corwin, H.G.,
Buta, R.J., Paturel, G., \& Fouque, P. 1991 ``Third Reference Catalaogue
of Bright Galaxies'', Springer-Verlag: New York

\par\noindent
Freedman, W.~L. 1988, \apj, 326, 691

%\par\noindent
%Freedman, W.~L., {\it et al.} 2001, \apj, 553, 47

%\par\noindent Freedman, W.~L., Madore, B.F., Rigby, J., Persson, S.E.,
%\& Sturch, L. 2008, \apj, 679, 71

\par\noindent
Gieren W., et al. 2006, \apj, 647, 1056

\par\noindent
Grieve, G.R Madore, B.F., \& Welch, D.L. 1985, \apj, 294, 513

\par\noindent Jackson, D.C., Skillman, E.D., Gehrz, R.D., Polomski,
E., Woodward, C.E., 2007a \apj, 656, 818

\par\noindent Jackson, D.C., Skillman, E.D., Gehrz, R.D., Polomski,
E., Woodward, C.E., 2007b \apj, 667, 891

%\par\noindent
%Madore, B.~F., \& Freedman, W.~L. 1991, \pasp, 103, 933

\par\noindent Madore, B.F., Freedman, W.~L., Rigby, J., Persson, S.E.,
\& Sturch, L. 2009a, \apj, (accepted)

\par\noindent Madore, B.F., Rigby, J., Freedman, W.~L., Persson, S.E.,
\& Sturch, L. 2009b, \apj, (submitted)

\par\noindent
McAlary, C.W., Madore, B.~F., \& Davis, L.E. 1983, \apj, 276, 487

\par\noindent Mantegazza, L., Antonello, E.,  Fugazza, D., Bossi, M., 
\& Covino, S.,  2001, \aap, 367, 759

\par\noindent
Pietrzynski, G., et al. 2006, \apj, 642, 216

\par\noindent Persson, S.E., Madore, B.F., Krzeminski, W., Freedman,
W.L., Roth, M., \& Murphy, D.C. 2004, \aj, 128, 2239

\par\noindent
Reach, W.~T., et al. 2005, \pasp, 117, 978

\par\noindent
Sandage, A.R. 1971, \apj, 166, 13

\par\noindent Schlegel, D.L., Finkbeiner, D.P., \& Davis, M. 1998,
\apj, 500, 525

\par\noindent
Udalski, A., Wyrzykowski, L., Pietrzynski, G., Szewczyk, O., Szymanski, M., 
Kubiak, M., Soszynski, I., \&  Zebrun, K. 2001, Acta Astron., 51, 221 
\par\noindent

\vskip 0.75cm

\vfill\eject

\noindent

\begin{deluxetable}{lcccc}
\tablecolumns{5}
\tablewidth{5.5truein}
\tablecaption{Mid-Infrared (IRAC) Magnitudes for Cepheids in IC~1613}
\tablehead{
\colhead{Cepheid}  & \colhead{~~~~~log(P)~~~~~}  & \colhead{3.6$\mu$m}    & \colhead{4.5$\mu$m}  & \colhead{8.0$\mu$m } 
\\ \colhead{}    & \colhead{(days)}  & \colhead{(mag)}   & \colhead{(mag)} &\colhead{(mag)} \cr
}
\startdata
IC~1613:[S71] V20   & 1.619 & 16.49 & 16.45 & 16.19  \\
      & & 0.09 & 0.17 & . . .  \\
IC~1613:[U01] 07647 & 1.219 & 17.56 & 17.87 & . . .  \\
      & & 0.09 & 0.17 & . . .  \\
IC~1613:[S71] V16   & 1.019 & 18.34 & 18.42 & . . .  \\
      & & 0.09 & 0.17 & . . .  \\
IC~1613:[S71] V06   & 0.973 & 18.68 & 18.64 & . . .  \\
      & & 0.08 & 0.16 & . . .  \\
IC~1613:[S71] V24   & 0.829 & 19.20 & 19.03 & . . .  \\
      & & 0.07 & 0.16 & . . .  \\
IC~1613:[S71] V22   & 2.093 & 15.29 & 15.24 & 14.95  \\
      & & 0.09 & 0.17 & . . .  \\
\enddata
\end{deluxetable}

\begin{deluxetable}{ccl}
\tablecolumns{3}
\tablewidth{5.0truein}
\tablecaption{Apparent (Cepheid) Moduli for IC~1613}
\tablehead{
\colhead{Bandpass}  & \colhead{Apparent Modulus}  & \colhead{Reference}
}
\startdata
 B& 24.56 (0.10) & Freedman (1988)\\
 V& 24.54 (0.09) & Udalski et al. (2001)\\
 R& 24.48 (0.11) & Freedman (1988)\\
 I& 24.44 (0.11) & Udalski et al. (2001)\\
 J& 24.39 (0.04) & Pietrzynski et al. (2006)\\
 H& 24.31 (0.09) & McAlary, Madore \& Davis (1984)\\
 K& 24.30 (0.05) & Pietrzynski et al. (2006)\\
 3.6$\mu$m& 24.29 (0.07) & this paper\\
 4.5$\mu$m& 24.28 (0.07) & this paper\\
\enddata
\end{deluxetable}

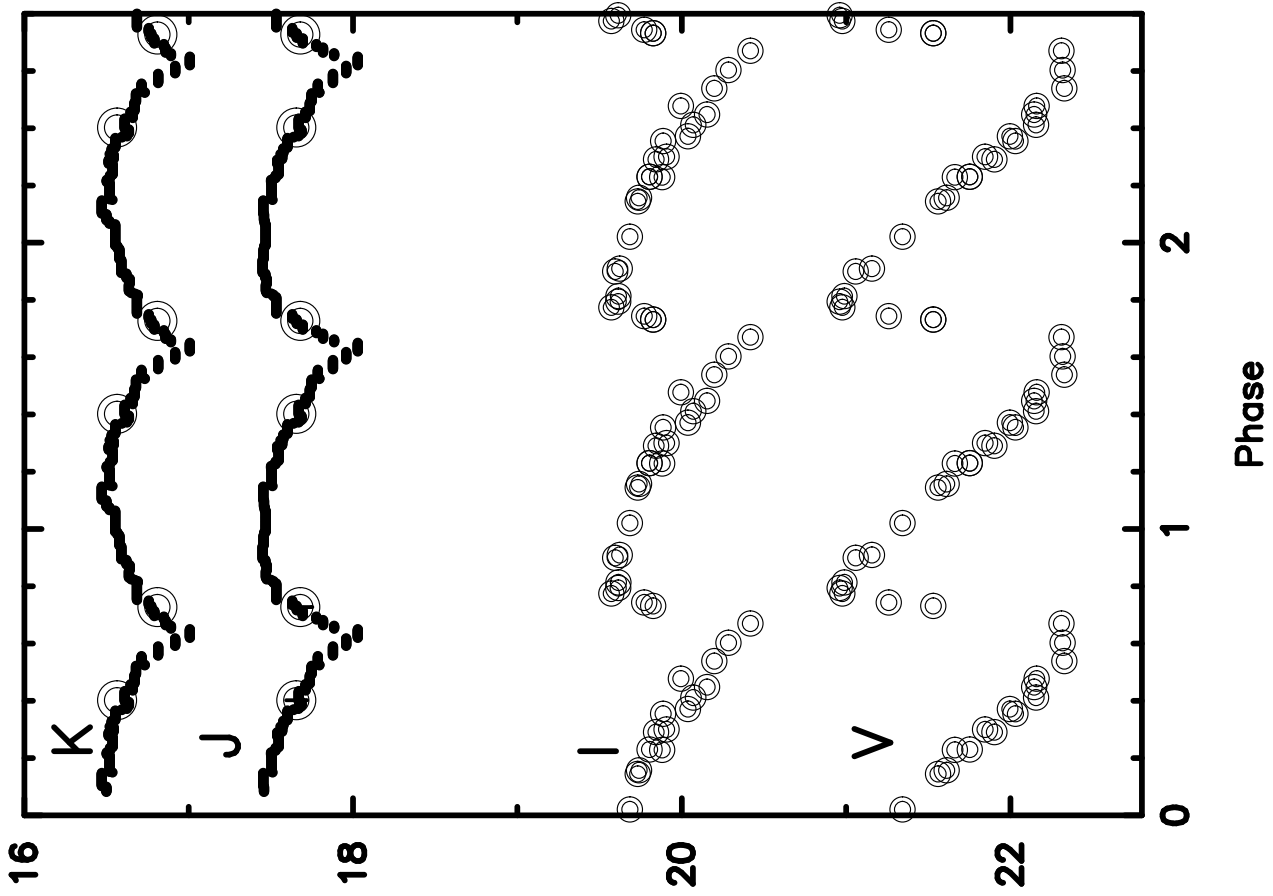
\end{document}